\documentclass[doublecol]{epl2}
\usepackage{amsmath}
\usepackage{xcolor}
\usepackage{amssymb}
\usepackage{graphicx}
\usepackage[colorlinks,bookmarks=false,citecolor=blue,linkcolor=blue,urlcolor=blue]{hyperref}

\DeclareFontFamily{U}{wncy}{}
\DeclareFontShape{U}{wncy}{m}{n}{<->wncyr10}{}
\DeclareSymbolFont{mcy}{U}{wncy}{m}{n}
\DeclareMathSymbol{\Sh}{\mathord}{mcy}{"58} 

\title{Superfluidity vs prethermalisation in a nonlinear Floquet system}

\author{S. Mu \inst{1} \and N. Mac\'e\inst{3} \and J. Gong \inst{1,2}\and C. Miniatura \inst{2,4} \and G. Lemari\'e\inst{2,3,4} \and M. Albert\inst{5}}

\shortauthor{S. Mu \etal}

\institute{
  \inst{1} Department of Physics, National University of Singapore, Singapore 117542, Singapore\\
  \inst{2} Centre for Quantum Technologies, National University of Singapore,Singapore 117543, Singapore\\
  \inst{3} Laboratoire de Physique Th\'eorique, IRSAMC, Universit\'e de Toulouse, CNRS, UPS, France\\
  \inst{4} MajuLab, International Joint Research Unit UMI 3654, CNRS, Universit\'e C\^ote d'Azur, Sorbonne Universit\'e, National University of Singapore, Nanyang Technological University, Singapore\\
  \inst{5} Universit\'e C\^ote d'Azur, CNRS, INPHYNI, France
}

\abstract{
  We show that superfluidity can be used to prevent thermalisation in a nonlinear Floquet system. Generically, periodic driving boils an interacting system to a featureless infinite temperature state. Fast driving is a known strategy to postpone Floquet heating with a large but always finite boiling time. In contrast, using a nonlinear periodically-driven system on a lattice, we show the existence of a continuous class of initial states which do not thermalise at all. This absence of thermalisation is associated to the existence and persistence of a stable superflow motion.
  }

\begin{document}

\maketitle

\section{Introduction}

Thermalisation of isolated quantum systems has been the subject of intensive research during the past decade. In this context, it is important to distinguish ergodic systems, who dynamically explore all possible energetically accessible states, from non-ergodic ones who can only explore a limited fraction of such states. Beside fine-tuned integrable systems whose dynamics is constrained by constants of motion, a seminal example of a nontrivial non-ergodic behavior is many-body localisation happening in strongly disordered quantum interacting systems~\cite{Huse2015,Abanin2019}. Other notable ergodicity breaking phenomena include quantum many-body scars and Hilbert space fragmentation~\cite{Serbyn2021,Regnault2021}.

In conservative systems, the dynamics can only take place within the energy shell since energy is conserved. In time-dependent systems, this is no longer true and time-periodic driving has been used to realize new non-equilibrium phases of matter, e.g. Floquet topological insulator and discrete time crystal~\cite{Weitenberg2021,Else_prx2017}. Generically, periodically-driven systems subject to interaction tend to evolve toward a featureless state akin to infinite temperature at long times~\cite{Luca_prx2014,PONTE2015196,Reitter_prl2017}. Different strategies have been devised to escape or hamper this heating mechanism due to energy exchanges. A first example is achieving many-body localisation through strong disorder in quantum systems~\cite{Ponte_prl2015,Moessner_prl2015,Khemani_prl2016,Else_prl2016}. Another example is to bring the system into a long-lived prethermalization state where Floquet energy transfers are largely suppressed and heating is postponed to exponentially large times. This prethermal state can be achieved by driving the system sufficiently fast. This mechanism is interesting because it is universal as it applies to both quantum~\cite{Abanin_prl2015,Bukov2015,Mori_prl2016,Mallayya_prx2019,Bloch_prx2020,Peng2021,Sagi2022} and classical systems~\cite{Mori_prb2018,Rajak_prb2019,Howell_prl2019,Hodson_prr2021} and does not rely on strong disorder. Nevertheless fast driving does not prevent ultimately the system to boil to an infinite temperature state.

In this Letter, we investigate the robustness of superfluidity in interacting Floquet systems as it could offer an escape route against prethermalisation. Indeed, in conservative systems, a superfluid at zero temperature \cite{leggett1999} is immune to small enough perturbations. As long as its velocity remains smaller than a well defined critical velocity \cite{Landau1941}, scattering by impurities is suppressed and internal excitations cannot be activated, preventing energy redistribution and thus thermalisation. It is only above this threshold that the system may enter a route toward wave thermalisation \cite{WT_theory1} and eventually reach a statistical equilibrium with an effective temperature. First discovered in liquid Helium \cite{allen38,kapitsa38}, superfluidity was later shown to be more universal and was observed in various quantum fluids \cite{PhysRevLett.28.885, PhysRevLett.83.2502,Amo2009,michel2018}.

Recent studies show that quenched superfluids generally evolve to a thermal equilibrium state at large times, see e.g. \cite{Buchhold2016pra, scoquart2022}. In this Letter, we show that, under suitable conditions, superfluidity can be maintained at all times in Floquet systems and escape prethermalisation and heating. Using a lattice model with periodically-kicked nonlinear interaction and onsite potentials, we build a driving protocol allowing the system to remain superfluid despite energy injection from the kicks, and show that the time $\tau_{boil}$ needed to boil the system diverges at the prethermal to superfluid transition.

Our paper is organized as follows. We first introduce our periodically-driven system and the new driving protocol allowing for superfluidity to emerge dynamically. 
We discuss its mapping to, and difference with, conservative superfluids. Superfluidity and heating features are then characterized by gradually increasing the complexity of the model. In the clean case, we perform a Bogoliubov analysis to unveil two important characteristic scales, the sound velocity and the healing length. We also describe the Floquet heating instability due to the presence of a nonlinearity. We then study the effect of a single impurity. 
Two distinct phases are found, a superfluid and a prethermal phase, exhibiting dramatically different dynamical behaviours. A superfluid flow is observed up to a critical velocity given by a Landau criterion \cite{Landau1941,hakim97,leboeuf2001}. Above that, the system reaches a prethermal phase with a large but finite boiling time. 
Strikingly, we find that the boiling time diverges at the transition between the superfluid and prethermal regions (up to the longest times accessible numerically). Thus, thermalisation is absent in the superfluid phase. Finally, we show that this phenomenology remains valid in the presence of disorder. We describe the statistical properties of the critical velocity using extreme value statistics and we find a very good agreement with numerical results. Technical details are given in the supplementary material (SM).

\section{Model} 
Driven systems with periodically-kicked onsite potentials have attracted extensive attention to study the dynamical localization transition and engineer exotic topological phases of matter~\cite{Casati1992prl,Chen2014prb,Sacramento2017prb,gong2018prl}. There has also been a rising interest with temporally modulated interactions to design synthetic gauge fields or modify the transport properties~\cite{gong2009prl1,gong2009prl2,Greschner2014prl,Meinert2016prl,Cherroret2022pra}. In the same spirit, we consider here bosonic particles hopping in a one-dimensional lattice (with unit lattice constant $a=1$) comprising $N$ sites (with periodic boundary conditions) where both the onsite potential and the mean-field (repulsive) interaction terms are periodically-kicked with the same time sequence (time period $T=1$ set to unity, driving frequency $\Omega = 2\pi$). Setting $\hbar=1$, the dynamics of our system is then governed by the following time-dependent Gross-Pitaevskii (GP) lattice equation
\begin{equation}\label{eq:KRGP}
\mathrm{i}\partial_t\psi_x= -\frac{J}{2}(\psi_{x+1}+\psi_{x-1}) + \Sh (t) \, (V_x + \tilde{g} N_a\vert \psi_x\vert^2)\psi_{x},
\end{equation}
where $\Sh (t)=\sum_{n\in\mathbb Z}\delta(t-n)$ is the Dirac comb, $x$ labels the lattice sites, $J>0$ is the nearest-neighbor hopping amplitude, $V_x$ the onsite potential, $N_a$ the number of particles and $\tilde{g} >0$ the two-body interaction strength. The wave function is normalized to $\sum_{x=1}^N|\psi_x(t)|^2 = 1$. In the following, we will consider $V_x = 0$ (clean case), $V_x = W \delta_{x,x_0}$ (single-site impurity with strength $W$ located at some given site $x_0$), and $V_x\in [-W/2,W/2]$ (site-uncorrelated uniformly distributed random potential). For future purposes, we define the renormalised nonlinear interaction strength $g=\tilde{g} N_a/N$. This model system could possibly be realized with cold atoms hopping in an optical lattice in the tight-binding regime. An additional optical potential (disordered or not) would be periodically flashed on the atoms. The interaction could also possibly be repeatedly switched on and off using a Feshbach resonance~\cite{Cheng2010rmp}.

Noteworthy, the system described by Eq.~\eqref{eq:KRGP} is formally equivalent to a nonlinear quantum kicked rotor (QKR) \cite{QKR_GPE1}. The QKR is a driven system exhibiting a rich phenomenology, from quantum chaos~\cite{Izrailev1990,Haake2006} to topological effects~\cite{beenakker2011,gong2016}. In particular, the QKR displays dynamical localisation~\cite{Casati1979}, a phenomenon analogous to Anderson localisation~\cite{Abrahams2010} but in momentum space~\cite{Fishman1984}. This system has been successfully implemented with cold atoms, see e.g.~\cite{Raizen1994,Garreau2008,Delande2009}. A variant of the QKR includes mean field GP interactions~\cite{QKR_GPE1,QKR_GPE2,Garreau2020} and has been used to study the breakdown of dynamical localization~\cite{QKR_GPE1,cherroret2014,Weld2021,Gupta2021} and more recently prethermalization and wave-condensation~\cite{Haldar2021}.

Importantly, we consider here a driving protocol where the potential $V_x$ is turned on adiabatically instead of abruptly. Sudden quench effects have been recently studied with the nonlinear QKR model in the regime of strong nonlinearities. It has been found that the system shows interesting prethermal properties before reaching an infinite-temperature thermal behavior at large times~\cite{martinez2022}. By contrast, we focus on an adiabatic driving protocol where an additional controlling parameter $a_t$ is coupled to $V_x$, whose strength is slowly ramped up as $a_t = \tanh(t/\tau)$ with $t\in \mathbb{Z}$ and $\tau=10^3$ unless specified otherwise. We will consider the subsequent dynamics at times $t \gg \tau$ ($a_t \sim 1$) where the onsite potential has reached its stationary value. Defining $\psi_x^{n^\pm}=\psi_x(t=n+0^\pm)$ and $\psi_k^{n^\pm}=\psi_k(t=n+0^\pm)$ with $\psi_k^n=\sum_x\psi_x^ne^{ikx}$ where $k\in[-\pi,\pi)$ is the quasi-momentum, the dynamics of our system is obtained by iterating the following nonlinear map, 
\begin{equation}
\label{eq:Our_map}
\begin{gathered}
\begin{aligned}
\psi_k^{n+1^-} &= e^{\mathrm{i} J \cos k} \, \psi_k^{n^+},\\
\psi_x^{n+1^+} &= e^{-\mathrm{i} a_n V_x} \, e^{-\mathrm{i} g\vert \psi_x^{n+1^-}\vert^2} \, \psi_x^{n+1^-}. 
\end{aligned}
\end{gathered}
\end{equation}

At this stage, it is interesting to comment on the connection of our model with the well-known conservative case. Our model Eq.~\eqref{eq:Our_map} is explicitly time-dependent, and we study its dynamics at stroboscopic times. This gives rise to the existence of Floquet quasi-energy bands. Like in \cite{Haldar2021}, we will consider a situation where these quasi-energy bands are well separated from each other so that their coupling, due to interactions, is weak. This implies constraints on the parameters of the model which we will discuss later.
In this regime, it is also possible to work in the low quasi-energy sector, i.e. at the edge of a quasi-energy band. This is achieved by considering initial plane wave states with sufficiently small quasi-momenta, see \cite{Haldar2021}. Moreover, Floquet heating due to the interplay between periodic driving and interactions must be considered in our model. Indeed, since energy is not conserved, the system is generally expected to evolve toward a featureless state which maximizes the entropy akin to infinite temperature at long times.

\section{Superfluid properties in the clean case} We will now show that our model shares some important formal features with the usual conservative superfluids, in particular the concepts of sound velocity and healing length. Adapting Bogoliubov theory to our periodically-driven system, we derive a low-energy excitation spectrum with a linear dispersion relation at low momenta that supports sound waves. Along this analysis, a threshold on the hopping amplitude $J$ is obtained to avoid fast Floquet heating~\cite{Lellouch2017, Haldar2021}.

Superfluidity can be characterized by a stability analysis of the initial plane-wave mode, $\psi_x(t=0) = \frac{1}{\sqrt{N}}e^{-\mathrm{i}k_0x}$. In the clean situation, the onsite potential is zero and the plane wave $\psi_x^0(t)=\frac{1}{\sqrt{N}}e^{-\mathrm{i}\phi(t)}e^{-\mathrm{i}k_0x}$ is an exact solution when $\dot\phi(t)=g \, \Sh (t)-J\cos k_0$. In order to obtain the low-energy excitation spectrum, we perform a linear stability analysis with the ansatz $\psi_x(t)=\psi_x^0(t)(1+\delta\psi_x(t))$. We decompose the perturbation $\delta\psi_x(t)=\sum_q u(q,t)e^{-\mathrm{i}qx}+v^*(q,t)e^{\mathrm{i}qx}$ into different plane-wave modes labeled by $q\in [-\pi,\pi)$. Linearizing Eq.~\ref{eq:KRGP}, we obtain the time-dependent Bogoliubov-de Gennes equation for the perturbation:  
\begin{equation}
i \partial_{t}
\begin{pmatrix}
u\\
v
\end{pmatrix}
=
\mathcal{M}(q,t) 
\begin{pmatrix}
u\\
v
\end{pmatrix}.
\end{equation}
Introducing $\lambda =2J\sin(\frac{q}{2})\sin(\frac{q}{2}+k_0)$ and the Pauli matrices, we have $\mathcal{M}(q,t) = \lambda \, \sigma_z + g \Sh(t) \, (\sigma_z+i \sigma_y)$. Since the operator $\mathcal{M}(q,t)$ is periodic in time, we consider the one-period evolution operator associated with it, aka the Floquet operator, given by the time-ordered integration $U(q) = \mathcal{T}e^{-i\int_0^1 dt \mathcal{M}(q,t)}$, that is
\begin{eqnarray}
\label{floquet_op}
U(q) =
e^{-\mathrm{i} \lambda \, \sigma_z} \ 
e^{-\mathrm{i}g \, (\sigma_z+i \sigma_y)} \; .
\end{eqnarray}

We next employ the Baker-Campbell-Hausdorff formula~\cite{Achilles2012bch} to approximate the effective Floquet Hamiltonian for $U(q)$ order by order in both $g$ and $\lambda$, i.e. $H_F(q)=-\mathrm{i}\ln U(q)\approx H^{(1)}_F(q)+H^{(2)}_F(q)+\cdots$. At first order in $g$ and $\lambda$, we find
\begin{eqnarray}
H^{(1)}_F(q) = (\lambda +g) \, \sigma_z + i g \, \sigma_y
\end{eqnarray}
with eigenvalues $\omega^{(1)}_{\pm}(q) = \pm\omega(q)$ where $\omega(q) = \sqrt{(\lambda + g)^2-g^2}$. At $k_0=0$, we get:
\begin{equation}\label{bog_spectrum}
  \omega(q)= 2\sqrt{gJ} \, \left|\sin \frac{q}{2}\right| \ \sqrt{\frac{J}{g}\sin^2 \frac{q}{2} + 1}.
\end{equation}
At small momenta, $\sin(q/2) \sim q/2$ and this excitation spectrum adopts a form similar to the Bogoliubov spectrum for time-independent systems with a quadratic kinetic energy term with particle mass $m\sim 1/J$. In the limit $q\to 0$, we recover a linear dispersion relation $\omega(q) \sim c q$ where $c=\sqrt{gJ}$ plays the role of a sound velocity. We also see from the square-root term in Eq.\eqref{bog_spectrum} that $\xi=\sqrt{J/g}$ is a length scale playing the same role as the healing length in usual superfluid systems.

\begin{figure}
\includegraphics[width=0.99\columnwidth]{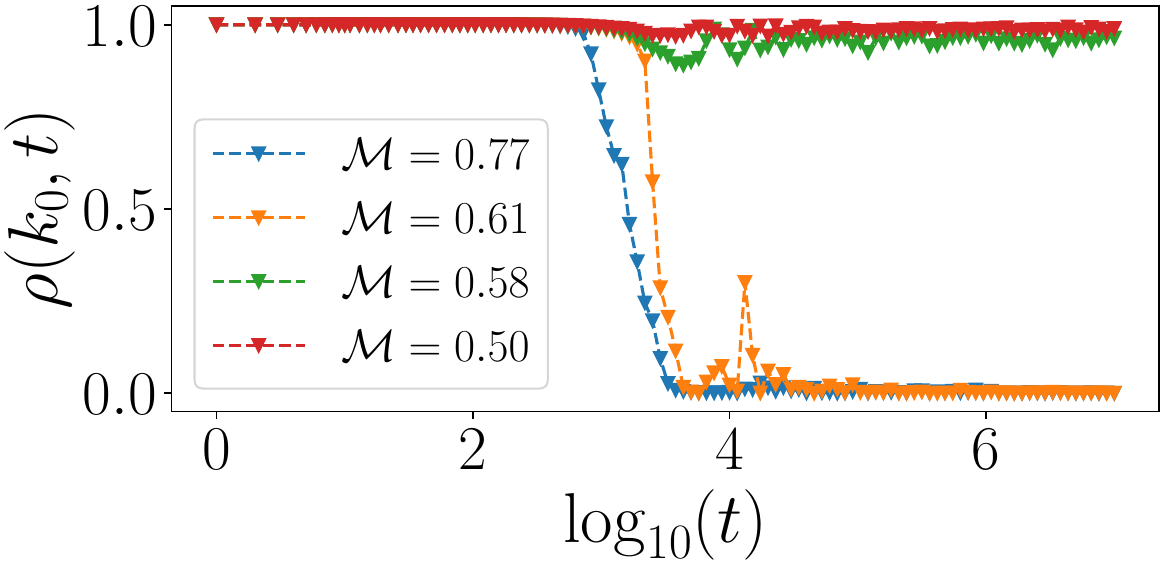}
\caption{Stroboscopic temporal dynamics of the population $\rho(k_0,t)$ of the initial mode for different initial momenta $k_0$ (Mach number $\mathcal M = k_0\xi$ where $\xi$ is the healing length). System parameters values are $J=1.3$, $g=0.1$ ($\xi \approx 3.6 $, $c \approx 0.36$), $\Lambda=W/c=0.41$ and $N=1024$. This population remains almost unchanged when $k_0$ is below a certain critical value while a rapid drop of the population takes place when $k_0$ is above that critical value.}
\label{fig:rho_dynamica}
\end{figure}

At this level of approximation, the excitation dynamics of our periodically driven system is identical to that for a time-independent system described by $H^{(1)}_F(q)$. In particular, Floquet heating is totally absent. It will arise when taking into account higher order terms in the expansion of $H_F(q)$. As shown in the SM, it is possible to exactly diagonalise $U(q)$ in Eq.~\eqref{floquet_op} and look for dynamical instabilities (eigenvalues of the Floquet operator outside the unit disk in the complex plane). At $k_0=0$, the first unstable mode is $q=\pi$ and appears when $J+g>\pi/2$ which defines a threshold for Floquet heating. Actually, this criterion has a simple interpretation when considering the relation between the driving frequency $\Omega=2\pi$ and the effective bandwidth $E_b$ of our model~\cite{Bukov2015,Haldar2021,Rosch2015}. The single-particle bandwidth is given by $E_b=2J$, and the correction from non-linearity results in $E_b=2(J+g)$ for small $g$. An estimate for suppressing the direct inter Floquet band transition is achieved by setting $E_b=2(J+g)<\pi$ to avoid fast Floquet heating. In the following, we will set the hopping amplitude $J<(\pi/2-g)$ and $g=0.1$.

\section{Superfluid flow across a single impurity} Here, we investigate further the analogy between our periodically-driven system and a conservative superfluid by studying the fate of an initial plane wave in the presence of a single impurity of strength $W$ located at some given site $x_0$, $V_x=W\delta_{xx_0}$, that is ramped up adiabatically to avoid sudden quench effects. In the following, since we work with periodic boundary conditions, we set $x_0=N/2$ without any loss of generality. We show that, below a certain critical velocity threshold, the system maintains its superfluid properties, i.e. can flow through the impurity without any dissipation or back scattering. 

In traditional superfluids, the nature of the flow interacting with a single localized impurity is governed by two independent dimensionless parameters (see \cite{hakim97,leboeuf2001} and SM). The first one is the ratio between the velocity of the fluid and the sound velocity, {\it aka} the Mach number $\mathcal M$. The second parameter is the dimensionless impurity strength $\Lambda$ (see SM). In this ($\mathcal M, \Lambda$) parameter space, there is a critical line that separates the superfluid phase existing at low Mach numbers from a non-superfluid phase existing at higher Mach numbers. In the superfluid regime, the impurity cannot induce any excitation in the fluid and the flow is stationary. It therefore lasts forever. Above the critical line, the flow is slowed down by energy transfers from the coherent motion to internal excitations and enters a weak turbulent regime characterized by erratic and complex dynamics.

Such a critical line also exists in our periodically-driven system. In our case, since $J\sim 1/m$ and $\hbar=1$, the velocity corresponds to $Jk_0$ and the Mach number simply reads $\mathcal M=Jk_0/c = k_0\xi$ whereas the effective impurity strength is given by $\Lambda=W\xi/J = W/c$, see Table $1$ in the SM. Concretely, we initialize the system in the single plane wave mode $\psi_x(t=0) = \frac{1}{\sqrt{N}}e^{-ik_0x}$ with quasi-momentum $k_0$. We then iterate the quantum map Eq.~\ref{eq:Our_map} with $V_x=W\delta_{x,x_0}$ for a given $W$. The nature of the flow is then established by looking at the evolution of the population of the initial mode $\rho(k_0,t)=|\psi_{k_0}(t)|^2$ over a very long time. Figure \ref{fig:rho_dynamica} illustrates our numerical experiment for a given set of parameters. Remarkably, $\rho(k_0,t\gg\tau)$ shows two different behaviors. For initial momenta $k_0$ below a certain critical threshold $k_c$, the population of the initial plane wave remains very close to unity, meaning that a superflow is preserved. Above $k_c$, $\rho(k_0,t)$ drops to zero after a certain time, meaning that the initial state is only marginally populated compared to other modes and the superflow is lost.

\begin{figure}
\includegraphics[width=1\columnwidth]{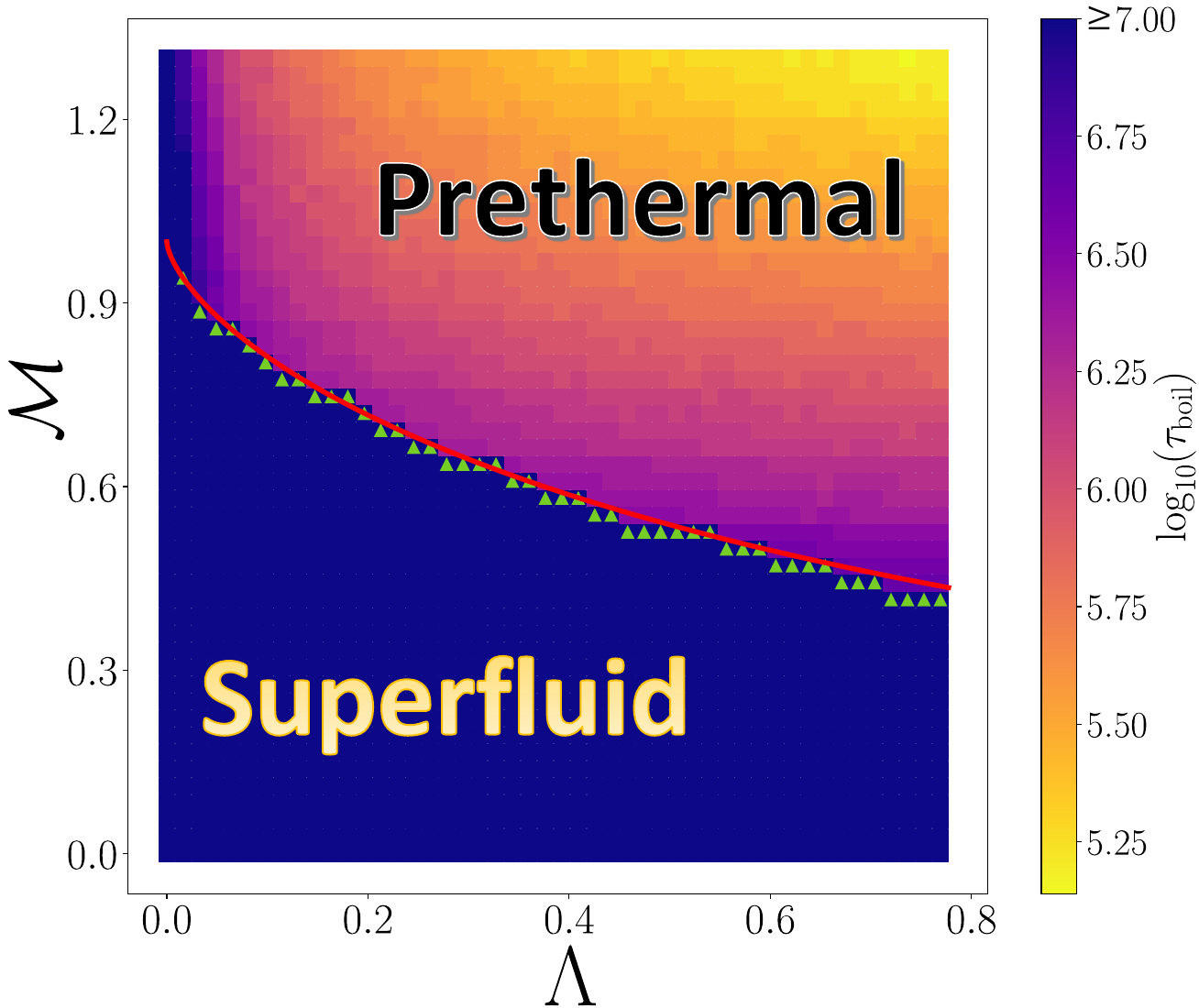}
\caption{Phase diagram of the fluid in the presence of a single-site impurity. In our case, the Mach number and renormalized impurity strength are $\mathcal M = k_0\xi$ and $\Lambda = W/c$, where $\xi =\sqrt{J/g}$ is the healing length and $c=\sqrt{gJ}$ is the sound velocity. The critical line (green triangles) delineating the superfluid state from a prethermal state is numerically determined from the time evolution of the population in the initial mode. Our data match remarkably well the analytical prediction Eq.\eqref{eq:vc_delta}, obtained for conservative quantum fluids (red solid line). In each region, the color map quantifies the boiling time $\tau_{\rm boil}$. Parameters values are set to $J=1.3, g=0.1$ and $N=1024$.}
\label{fig:single_imp}
\end{figure}

Next, we repeat the numerical simulation for different impurity strength $W$ to quantitatively extract the critical line in the ($\mathcal M,\Lambda$) plane, see Fig.\ref{fig:single_imp}. The agreement between our data, obtained for a Floquet system, and the analytical prediction for this critical line in a conservative quantum fluid~\cite{hakim97,leboeuf2001}, namely
\begin{equation}
  \Lambda_c(\mathcal M)=\frac{\sqrt{2}}{4\mathcal M}\sqrt{-8\mathcal M^4-20 \mathcal M^2+1+(1+8 \mathcal M^2)^{3/2}},
  \label{eq:vc_delta}
\end{equation}
is truly remarkable.

\section{Superfluid to prethermal transition} 


Above the critical velocity, the initial state is unstable and the flow enters a regime analogous to the weak turbulent regime in time-independent systems \cite{hakim97,leboeuf2001}. Following \cite{Haldar2021}, we expect our system to enter a long-lived prethermal state before Floquet heating drives the system into an infinite temperature state after some boiling time $\tau_{\rm boil}$. 
A natural question that then arises is whether, below the critical line, $\tau_{\rm boil}$ is finite (albeit larger than the observation time considered in Figure \ref{fig:rho_dynamica}), in which case, we would have a metastable superfluid state, or infinite, in which case we would have a true superfluid state. In this Section, we show that superfluidity prevents thermalisation at all times in our periodically-driven nonlinear system.

\begin{figure}
\includegraphics[width=.95\columnwidth]{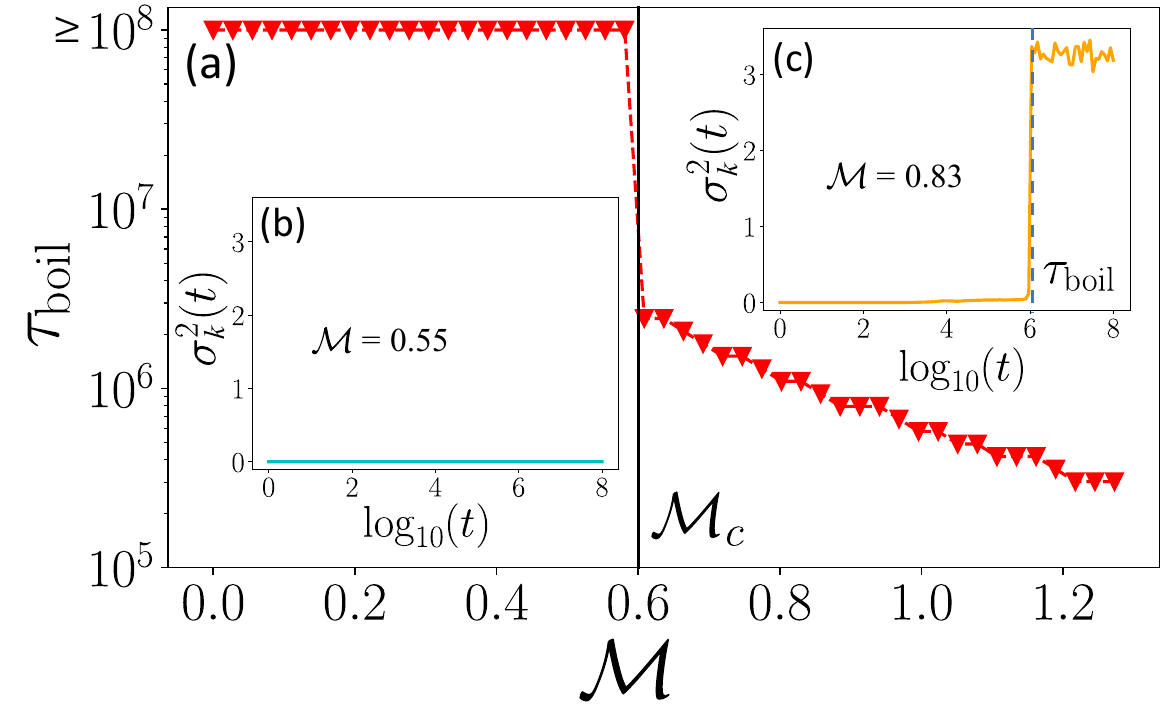}
\caption{Temporal behavior of the variance $\sigma_k^2(t)$ of the momentum distribution and boiling time obtained by iterating the nonlinear map Eq.~\ref{eq:Our_map} for different initial momenta $k_0$. Parameter values are $J=1.3$,  $g=0.1$, $\Lambda=W/c=0.39$ and $N=1024$. (a) Boiling time $\tau_{\rm boil}$ as a function of the Mach number $\mathcal M=k_0\xi$. The vertical black line marks the critical Mach number $\mathcal M_c$ inferred from the theoretical prediction Eq.~\eqref{eq:vc_delta}. (b) The variance for $k_0<k_c$ remains small and does not change (up to $10^8$ periods). (c) The variance for $k_0>k_c$ remains small for an exponentially long time before it abruptly jumps and saturates to $\pi^2/3$. The blue vertical dashed line defines $\tau_{\rm boil}$.}
\label{fig:tau_boil}
\end{figure}

The boiling process induced by Floquet heating can be conveniently characterised by the temporal evolution of the variance $\sigma_k^2(t)=\langle k^2\rangle-\langle k\rangle^2$ of the quasi-momentum distribution~\cite{Haldar2021}. In the prethermal regime $t \ll \tau_{\rm boil}$, $\sigma^2_k(t)$ shows a plateau while at $t\gg \tau_{\rm boil}$ it saturates to the value $\pi^2/3$ obtained for a uniform momentum distribution, a signature of an infinite temperature state. The transition from these two behaviors is sharp and the time at which this happens defines the boiling time. Figures~\ref{fig:tau_boil}(b) and (c) present two examples of simulations in, respectively, the superfluid regime and the prethermal regime. In the first case, the variance remains small and no boiling time could be defined up to the maximal integration time, i.e. $10^8$ periods. On the contrary, inset (c) demonstrates the existence of a finite boiling time although it is exponentially large. Note that, though the variance remains small before $\tau_{\rm boil}$ in both insets (b) and (c), the underlying dynamical behaviors are different, see SM.

We now present one of our most important results. In Fig.\ref{fig:tau_boil}a, we plot the evolution of $\tau_{\rm boil}$ as a function of the Mach number $\mathcal M=k_0\xi$ at fixed dimensionless impurity strength $\Lambda=W/c$. Strikingly, we find that $\tau_{\rm boil}$ jumps discontinuously when $\mathcal M$ crosses the critical value $\mathcal M_c$ for the superfluid regime previously determined in Figure \ref{fig:single_imp}. In the prethermal phase, $\tau_{\rm boil}$ is always finite while in the superfluid phase, thermalisation of the system does not take place up to the longest times considered ($t=10^8$). In Fig.\ref{fig:tau_boil}a, the jump in size of $\tau_{\rm boil}$ just below and above $\mathcal M_c$ is of nearly two orders of magnitude. This sharp transition at $\mathcal M_c$ strongly suggests that the superfluid is immune to Floquet heating. This is reflected in Fig.\ref{fig:single_imp} where the color map in the phase diagram represents $\tau_{\rm boil}$ as a function of $\mathcal M$ and $\Lambda$, as obtained from iterating the Floquet quantum map up to $10^7$ periods.  The excellent agreement between the critical line below which superfluidity is maintained and the critical line below which $\tau_{\rm boil}$ diverges leads us to conclude that a superfluid state below its critical velocity is stable in our periodically-driven system, i.e. robust against thermalisation due to either interaction with a localized impurity or to Floquet heating, or to their combined effects.

\section{Disordered case}

\begin{figure}
\centering\includegraphics[width=.99\columnwidth]{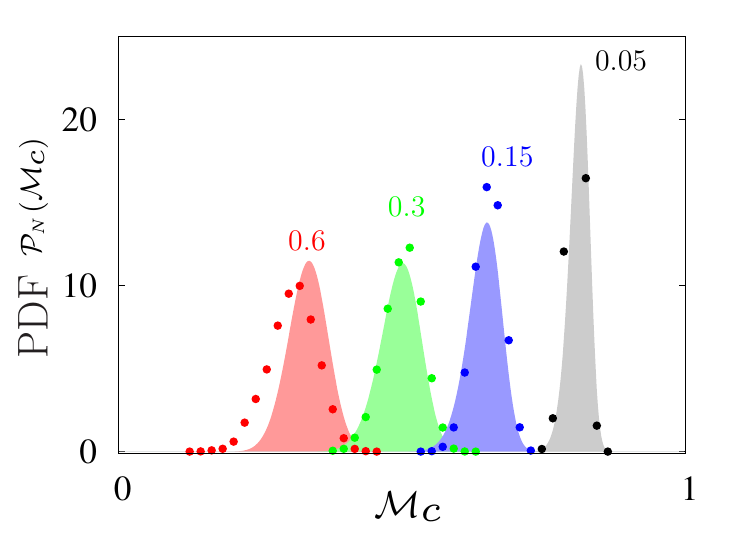}
\caption{Probability distribution function (PDF) $\mathcal P_N(\mathcal M_c)$ of the critical Mach number $\mathcal{M}_c=Jk_c/c$ for 4 different disorder strengths $\Lambda=W/c$ at fixed system size $N=1024$, interaction strength $g=0.1$ and hopping parameter $J=1$. Symbols are the histograms extracted from numerical simulations using $10^4$ disorder configurations: $\Lambda = 0.05$ (black dots),  $\Lambda = 0.15$ (blue dots), $\Lambda = 0.3$ (green dots) and $\Lambda = 0.6$ (red dots). The shaded curves are the corresponding analytical predictions from Eq.~\eqref{eq:PhiN}.}
\label{fig:vc_Wi}
\end{figure}

\begin{figure}
\centering\includegraphics[width=.95\columnwidth]{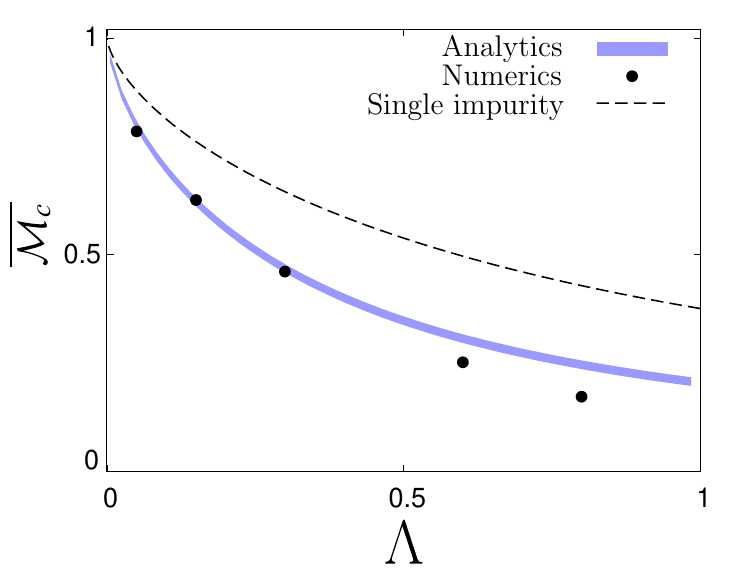}
\caption{Median critical Mach number $\overline{\mathcal M_c} = \overline{k_c}\xi$ as a function of $\Lambda=W/c$ for system size $N=1024$, interaction strength $g=0.1$ and hopping parameter $J=1$. Black dots: Data obtained from numerical simulations using $10^4$ disorder configurations. Thick blue line: Extreme value statistics prediction. Dashed line: Theory prediction for a single impurity of strength $W$.}
\label{fig:vc_median}
\end{figure}

We now address the dynamical behavior of our system in the presence of a site-uncorrelated disordered potential with uniform distribution $V_x\in[-W/2,W/2]$. As expected from \cite{albert2010}, our numerical investigations (data not shown) indicate that the phenomenology observed in Fig.\ref{fig:single_imp} and Fig.\ref{fig:tau_boil} with a single impurity remains entirely correct in the presence of a disordered potential. The only difference is that the critical line obtained in Fig.\ref{fig:single_imp} now depends on the configuration of disorder considered and $\mathcal M_c$ becomes a random variable. We therefore discuss here the statistical properties of this random critical momentum $\mathcal M_c$ as a function of the different parameters.

To compute the distribution of $\mathcal M_c$, we use arguments based on screening and extreme value statistics \cite{albert2010}. First, we recall that $\xi$ is the minimal length scale for density fluctuations. Therefore, spatial details of the disorder potential on scales below $\xi$ cannot be resolved by the system. We therefore renormalise the disorder through a coarse graining procedure over a scale of a few $\xi$. Then, we isolate the maximum value of the renormalised disorder and apply criterion Eq.(\ref{eq:vc_delta}) to obtain the critical velocity. Using extreme value statistics, it is possible to compute the full distribution of the critical velocity as it is explained in the SM. This approach gives a rather accurate description of the statistical properties of $\mathcal M_c$ as demonstrated below. Our central result is the cumulative distribution of the critical $\mathcal{M}_c$ which is given by the following formula
\begin{equation}
  \Phi_N(\mathcal{M}_c)=1-\left[\frac{1+\textrm{erf}\left(\frac{\Lambda_c(\mathcal{M}_c)}{\Lambda_0}\right)}{2}\right]^{\frac{N}{\alpha\xi}} 
  \label{eq:PhiN}
\end{equation}
with $\mathcal{M}_c=k_c\xi$, $\Lambda_0= \sqrt{\frac{\alpha\xi}{6}} \, \Lambda = \sqrt{\frac{\alpha\xi}{6}} \, \frac{W}{c}$ where $\alpha=2.78$ (see SM). Then the probability distribution function (PDF) of the critical $\mathcal{M}_c$ is simply the derivative of $\Phi_N$, $\mathcal P_N(\mathcal M)=\Phi'_N(\mathcal M)$. Note that this simple expression contains nontrivial scaling properties of $\mathcal{M}_c$ with respect to the system size $N$, the disorder strength $W$, the hopping amplitude $J$ and the non-linearity strength $g$ via $c$ and $\xi$. 

Fig.\ref{fig:vc_Wi} shows the comparison between numerically-extracted histograms of the critical $\mathcal{M}_c$ and our analytical predictions for different values of the disorder strength $W$. Our statistical model remarkably reproduces not only the typical critical $\mathcal{M}_c$ and its fluctuations but also the full probability distribution. In order to quantify the typical value of the critical $\mathcal{M}_c$, we have computed its median $\overline{\mathcal{M}_c}$, defined as $\Phi_N(\overline{\mathcal M_c})=1/2$, as a function of the disorder strength $W$ and the results are shown in Fig.\ref{fig:vc_median}. One can see that it is substantially below the critical Mach number predicted for a single impurity but very well captured by the effective disorder computed from extreme value arguments. In addition, we have also checked the scaling with the system size $N$ (see SM) and interaction strength $g$ (data not shown) which give the same level of agreement. Altogether, these results demonstrate that the superfluid to prethermal transition in the disordered case can be understood and accurately described by a coarse-graining procedure. In other words, the single-site impurity case can be seen as an effective description of the uniformly distributed disorder case provided its strength is appropriately renormalised. 

Last, we have also addressed the Floquet heating instability with disorder. We find that the distribution of the critical $\mathcal{M}_c$ determined above predicts very well the onset of a diverging $\tau_{\rm boil}$ (see SM). For a superfluid moving through disorder with initial $\mathcal{M}$ below its renormalised critical $\mathcal{M}_c$, thermalisation is not seen to occur under periodic driving.

\section{Conclusion}
In this letter, we have shown that superfluidity can be used to prevent thermalisation in a nonlinear Floquet system. While fast driving is a well-known mechanism to postpone Floquet heating to exponentially long times, we have demonstrated a model where the boiling time diverges at the prethermal to superfluid transition. 
This promotes superfluidity as a new mechanism for ergodicity breaking in Floquet systems. 

This opens interesting avenues to study Floquet topological superfluids \cite{topoSF2014} and dynamical phase transitions \cite{dynTransi2018} such as the recently proposed dynamical Berezinkii-Kosterlitz-Thouless phase transition \cite{cherroret2022}.   

\acknowledgments
We thank P. E. Larr\'e for fruitful discussions. This study has been supported by the French National Research Agency (ANR) under projects COCOA ANR-17-CE30-0024, MANYLOK ANR-18-CE30-0017 and GLADYS ANR-19-CE30-0013, the EUR grant NanoX No. ANR-17-EURE-0009 in the framework of the “Programme des Investissements d'Avenir”, and by the Singapore Ministry of Education Academic Research Fund Tier I (WBS No. R-144-000-437-114). Computational resources were provided by the facilities of Calcul en Midi-Pyr\'en\'ees (CALMIP) and the National Supercomputing Centre (NSCC), Singapore.

\bibliographystyle{eplbib.bst}

\end{document}


\maketitle

This supplementary contains the following material arranged by sections:
\begin{enumerate}
\item We present the correspondence between the superfluid states obtained in our periodically-kicked Gross-Pitaevskii model and in the usual time-independent Gross-Pitaevskii equation.
\item We detail the derivation of the Bogoliubov expansion for our model in the clean case ($V_x=0$).
\item We detail the derivation of the effective impurity strength in the disordered Floquet model.
\end{enumerate}

\section{Mapping to the Gross-Pitaevskii equation}

The one-dimensional Gross-Pitaevskii equation describing the time evolution of the macroscopic wave function $\Phi(x,t)$ of a system of identical bosons of mass $m$ interacting with a single impurity of strength $W$ located at the origin reads:
\begin{equation}
  \label{GP_eq}
  \begin{split}
    \mathrm{i}\hbar \partial_t\Phi(x,t)=&-\frac{\hbar^2}{2m}\frac{\partial ^2}{\partial x^2}\Phi(x,t)+g|\Phi(x,t)|^2\Phi(x,t)\\
    &+W\delta(x)\Phi(x,t).
  \end{split}
  \end{equation}
The normalisation condition reads $\int dx \,  |\Phi(x,t)|^2 = N_0$, where $N_0$ is the total number of atoms, and we call $n_0$ the atomic number density.

Setting $\Phi(x,t)=\exp(-\mathrm{i}\mu t/\hbar) \phi(x)$, we now look for a stationary solution with the following boundary condition:
\begin{equation}
  \phi(x)\underset{|x|\to+\infty}{\to}\sqrt{n_0}\,e^{\mathrm{i}m v_0 x/\hbar}. 
\end{equation}
It describes a macroscopic stationary wave function with asymptotic density $n_0$ and velocity $v_0$. Since $\phi(x)$ solves
\begin{equation}
  \label{GPStat_eq}
    \mu \phi(x) = -\frac{\hbar^2}{2m}\phi''(x)+g|\phi(x)|^2\phi(x)+W\delta(x)\phi(x),
  \end{equation}
where $\phi''$ denotes the second space derivative, we see that the boundary condition imposes
\begin{equation}
  \mu=gn_0+\frac{1}{2}mv_0^2.
\end{equation}
We introduce the sound velocity $c$ and the healing length $\xi$ according to the standard formulae \cite{pethick2008bose}
\begin{equation}
  mc^2=gn_0\,, \quad \xi=\frac{\hbar}{\sqrt{mgn_0}}=\frac{\hbar}{mc}.
\end{equation}
Using the rescaled variables $z=x/\xi$ and $\psi=\phi/\sqrt{n_0}$, Eq.~\eqref{GPStat_eq} takes on the following universal form:
\begin{equation}
  \label{GP_eq_universal}
  \left[1+\frac{1}{2}\mathcal{M}^2\right]\psi(z)=-\frac{1}{2}\psi''(z)+\Lambda \delta(z) \psi(z)+|\psi(z)|^2 \psi(z)
\end{equation}
which depends on two dimensionless parameters, the Mach number $\mathcal M$ and the effective impurity strength $\Lambda$ given by:
\begin{equation}
  \label{sf_parameters}
  \mathcal{M}=\frac{v_0}{c}\,,\quad \Lambda=\frac{m}{\hbar^2}\,W\xi \, = \frac{W}{\hbar c}.
\end{equation}
The existence of the superfluid solution $\psi(z)$ is therefore governed by $\mathcal{M}$ and $\Lambda$ only as demonstrated in~\cite{hakim97,leboeuf2001} for instance.

\begin{table}
  \centering
\begin{tabular}{|c|c|}
  \hline
  Continuous GP  & Kicked GP lattice \\
  Model Eq.\eqref{GP_eq}  & Model Eq.(1) in the main text \\
  \hline
  $p$ & $k$ \\
  $m$ & $1/J$ \\
  $g n_0$ & $g$\\
  $W$ & $W$ \\
  $v_0$ & $J k_0$\\
  $c=\sqrt{gn_0/m}$ & $c=\sqrt{Jg}$\\
  $\xi=\hbar/mc$ & $\xi=\sqrt{J/g}$\\
  $\mathcal{M}=v_0/c$ & $\mathcal{M}=J k_0/c$ $= k_0\xi$\\
  $\Lambda={m}W\xi/{\hbar^2}$ & $\Lambda={W\xi}/{J}$ $=W/c$\\
  \hline
\end{tabular}
\caption{\label{table_mapping} Mapping obtained at low energy between the continuous conservative Gross-Pitaevskii model given by Eq.\eqref{GP_eq} and the non-linear Floquet lattice model given by Eq.(1) in the main text (with lattice constant $a=1$ and $\hbar=1$).}
\end{table}

Realizing that the discrete version of the kinetic energy term $-\frac{\hbar^2}{2m}\frac{\partial ^2}{\partial x^2}\Phi(x,t)$ writes $-\frac{\hbar^2}{2ma^2}\big[\Phi_{x+a}(t)+\Phi_{x-a}(t)-2 \Phi_{x}(t)\big]$ on a lattice of spatial period $a$, we see that the hopping amplitude in this discrete version is simply $\hbar^2/(ma^2)$. The same scaling is obviously found from the dispersion relation of $-J \cos(ka)$ obtained in our lattice model in the absence of kicks. For $ka \ll 1$, we get $-J \cos(ka) \approx -J (1-(ka)^2/2)$. With $p=\hbar k$, the effective mass of our discrete model would simply be $\hbar^2/(Ja^2)$. Setting $\hbar =1$ and $a=1$, we thus get the mapping $J \leftrightarrow 1/m$ between the continuous GP model given above by Eq.\eqref{GP_eq} and our kicked GP lattice model given by Eq.(1) in the main text. With this in mind, the formal mapping given in Table~\ref{table_mapping} immediately follows. Along with the quantum map associated to  Eq.(1) of the main text, we choose to spotlight the momentum variables here. As a consequence, hereafter, we express the Mach number as $\mathcal{M}=k_0\xi$ in our kicked system. We also choose to express the effective impurity strength as $\Lambda = W/c$.
One of the major results of this Letter is that a superfluid state can persist under the Floquet dynamics just like it can persist in continuous conservative systems.

\begin{figure}
\includegraphics[width=0.95\columnwidth]{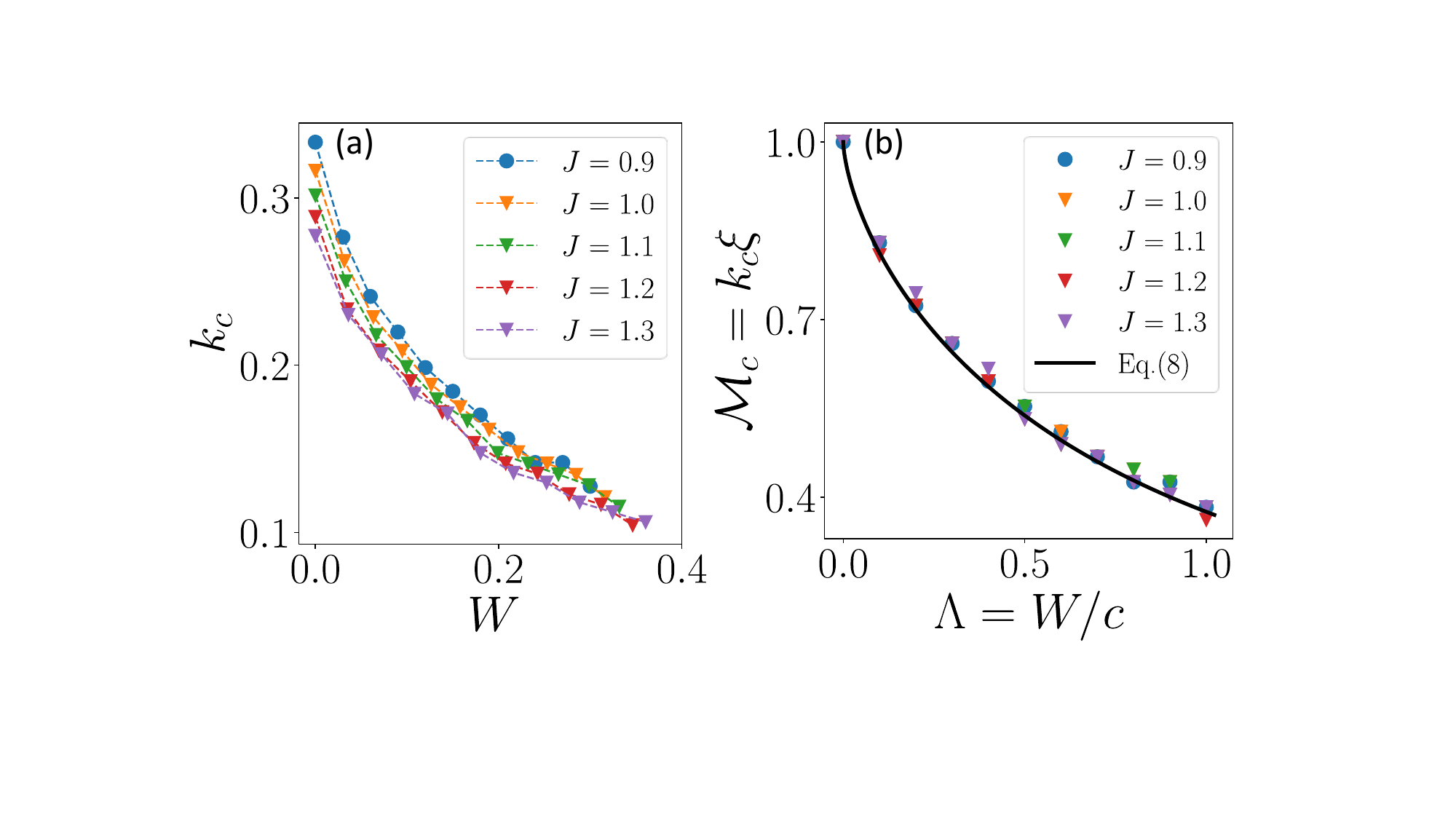}
\caption{(a) Numerically-extracted critical momentum $k_c$ (symbols) as a function of the bare impurity strength $W$ for different hopping amplitudes $J$ at interaction strength $g=0.1$ and system size $N=1024$. All curves are different. (b) When $k_c$ and $W$ are rescaled to $\mathcal M_c = k_0\xi$ and $\Lambda = W/c$, where $\xi$ is the healing length and $c$ the sound velocity, all previous curves collapse onto a single universal critical line. The analytical prediction (solid line) is given by $W/c=\Lambda_c(\mathcal M_c)$ through Eq.\eqref{eq:vc_delta}. As one can see, the agreement between the data (symbols) and the theory is very good.
}
\label{fig:imp_scaling_K}
\end{figure}

To complement the results on the critical Mach number $\mathcal M_c$ shown in Fig.2 of the main text, we present here additional results showing the scaling behavior of the critical curve when plotted in the ($\mathcal M,\Lambda$) plane. In Fig.\ref{fig:imp_scaling_K}a, we plot the numerically-extracted critical momenta $k_c$ as a function of $W$ for different hopping amplitudes $J$. As one can see, we obtain different non-overlapping curves. However when $k_c$ and $W$ are properly rescaled by $\xi$ and $c$, all these curves collapse onto a single universal critical line, see Fig.\ref{fig:imp_scaling_K}b. The agreement with the analytical prediction $W/c=\Lambda_c(\mathcal M_c)$, where
\begin{equation}
\label{eq:vc_delta}
\Lambda_c(y)=\frac{\sqrt{2}}{4y}\sqrt{-8y^4-20 y^2+1+(1+8 y^2)^{3/2}},
\end{equation}
proves just remarkable.

\section{Bogoliubov expansion in the periodically kicked system without impurities} 
In the main text, we have seen that a sound velocity $c=\sqrt{gJ}$ and a healing length $\xi=\sqrt{J/g}$ can be defined in our periodically-kicked system by looking at low-energy excitations on top of a stationary plane-wave solution. Here we provide more details about this derivation and give additional numerical results on the linear dispersion relation and on the dynamical instabilities.

Let us consider a clean superfluid ($V_x=0$) in a one dimensional lattice under periodic driving. Eq.(1) in the main text then reduces to
\begin{equation}\label{eq:clean_model}
\mathrm{i}\partial_t\psi_x= -\frac{J}{2}(\psi_{x+1}+\psi_{x-1})+ \tilde{g} N_a \, \Sh(t) \,\vert \psi_x\vert^2\psi_{x} ,
\end{equation}
where $\Sh(t) = \sum_{n\in\mathbb Z}\delta(t-n)$. We can readily check that any plane wave state $\psi_x^0(t)=\frac{1}{\sqrt{N}}e^{-\mathrm{i}\phi(t)}e^{-\mathrm{i}k_0x}$ is an exact solution of Eq.\eqref{eq:clean_model} provided $\dot\phi(t)=g\Sh (t)-J\cos k_0$ where $g=\tilde{g} N_a/N$. Given this time-dependent solution, the basic idea of the Bogoliubov expansion is to consider a small perturbation on top of it. There are several ways to parameterise the perturbation, and we choose the following form for technical convenience~\cite{Bukov2015}, $\psi_x(t)=\psi_x^0(t)(1+\delta\psi_x(t))$. 
Using this parametrisation and after linearising the nonlinear equation Eq.~\eqref{eq:clean_model}, we can obtain the time-dependent Bogoliubov-de Gennes equations for the perturbation, which take the general form,  
\begin{equation}
\label{eq:bdg_eqn}
\mathrm{i} \partial_{t}\left(\begin{array}{c}
\delta\psi_x(t) \\
\delta\psi^*_x(t)
\end{array}\right)=
\mathcal{L}(t)
\left(\begin{array}{c}
\delta\psi_x(t) \\
\delta\psi^*_x(t)
\end{array}\right),
\end{equation}
where
\begin{equation}
\mathcal{L}(t)=\left(\begin{array}{cc}
\mathcal{F}_x(t) & g\Sh (t)|\psi^0_x|^2\\
-g\Sh (t)|\psi^0_x|^2 & -\mathcal{F}^*_x(t)
\end{array}\right).
\end{equation}
Here, the linear operator $\mathcal{F}_x(t)$ acting on $\delta\psi_x(t)$ gives
\begin{eqnarray}
\mathcal{F}_x(t)\delta\psi_x(t) &=&-\frac{J}{2}\frac{\psi^0_{x+1}(t)\delta\psi_{x+1}(t)+\psi^0_{x-1}(t)\delta\psi_{x-1}(t)}{\psi^0_x(t)}\nonumber\\
&+& 2g\Sh (t)\delta\psi_x(t)-\frac{i\partial_t\psi^0_x(t)}{\psi^0_x(t)}\delta\psi_x(t).
\end{eqnarray} 
For the chosen exact solution, we have $\frac{i\partial_t\psi^0_x(t)}{\psi^0_x(t)}=\dot\phi(t)$ and $\psi^0_{x\pm 1}(t)/\psi^0_x(t) = \exp (\mp \mathrm{i} k_0)$. Next, we decompose the perturbation into different Fourier modes labeled by $q$,
\begin{equation}
\delta\psi_x(t)=\sum_qu(q,t)e^{-iqx}+v^*(q,t)e^{iqx}
\end{equation}
with $q \in [-\pi,\pi)$. Substituting the decomposition into Eq.~\ref{eq:bdg_eqn}, we have 
\begin{equation}\label{bogoliubov}
i \partial_{t}
\begin{pmatrix}
u\\
v
\end{pmatrix}
=
\mathcal{M}(q,t) 
\begin{pmatrix}
u\\
v
\end{pmatrix},
\end{equation}
where
\begin{equation}
\begin{gathered}
\mathcal{M}(q,t)= \lambda(q)
\begin{pmatrix}
1 & 0\\
0 & -1
\end{pmatrix}
+  g \, \Sh(t) 
\begin{pmatrix}
1 & 1\\
-1 & -1
\end{pmatrix} \\
= \lambda(q) \, \sigma_z + g \, \Sh(t) \, (\sigma_z+\mathrm{i}\sigma_y)
\end{gathered}
\end{equation}
with $\lambda(q)=2J\sin(\frac{q}{2})\sin(\frac{q}{2}+k_0)$ and the Pauli matrices $\sigma_y$ and $\sigma_z$.

\begin{figure}
\includegraphics[width=1.0\columnwidth]{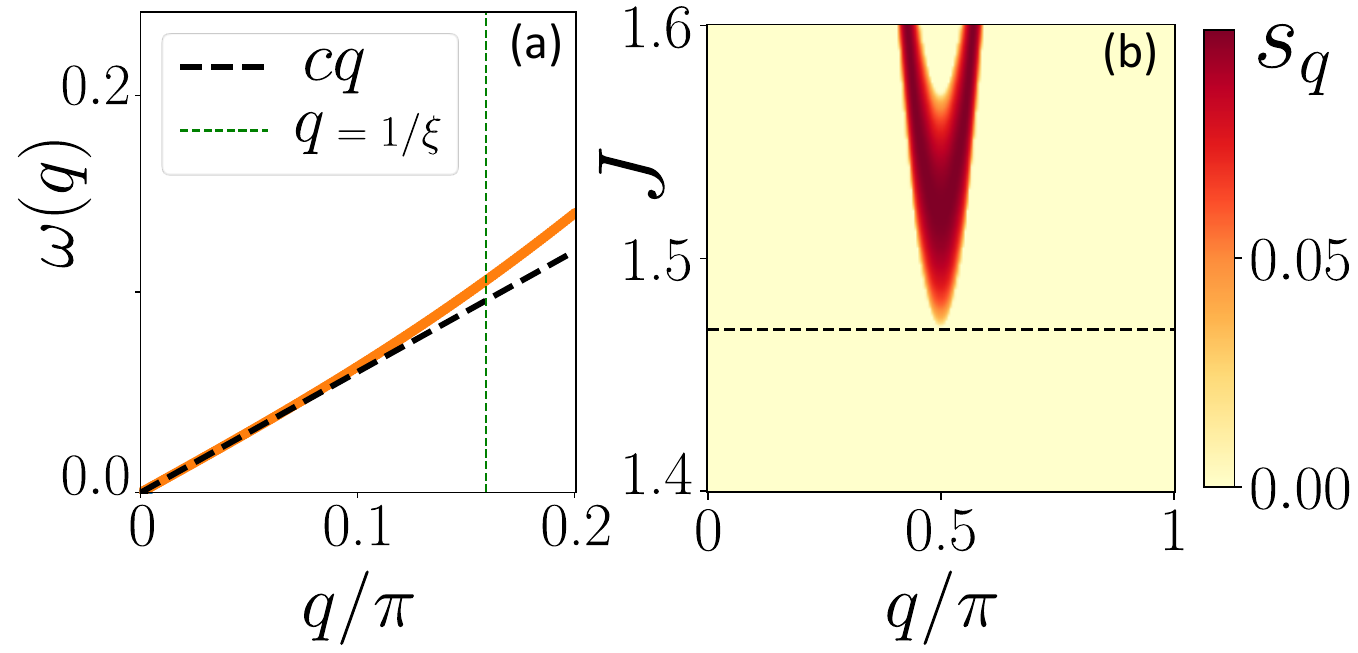}
\caption{(a) Quasi-energy $\omega(q)$ (positive branch) of the Floquet operator Eq.\eqref{floquet_op} as a function of $q$ (orange line). The linear dispersion relation with sound velocity $c=\sqrt{gJ}$ is indicated by the black dashed line. The green vertical dashed lines marks $1/\xi$ with $\xi=\sqrt{J/g}$. Parameters values are $k_0=0$, $g=0.1$ and $J=1$. (b) Dynamical instability diagram in the ($J,q$) plane. The instability rate $s_q=-{\rm Im}\omega(q)$ is quantified with the color map. Instability occurs at $q=\pi$ when $J+g\approx \pi/2$. The black dashed line denotes the instability threshold $J=\pi/2-g$. Parameters values are $k_0=0$ and $g=0.1$.
}
\label{fig:dispersion}
\end{figure}

The $1$-period Floquet operator associated to Eq.~\eqref{bogoliubov} reads:
\begin{eqnarray}
\label{floquet_op}
U(q) =
\mathcal{T}e^{-\mathrm{i}\int_0^1 dt \mathcal{M}(q,t)} 
= e^{-\mathrm{i}\lambda \, \sigma_z} \, 
e^{-\mathrm{i}g \, (\sigma_z+\mathrm{i}\sigma_y)}
\end{eqnarray}
where $\mathcal{T}$ denotes the time-ordered integration operator. It is a product of 2 non-commuting unitary operators. Next, we note that $(\sigma_z+\mathrm{i}\sigma_y)$ is a nilpotent matrix as it squares to zero. As a consequence, 
\begin{equation}
e^{-\mathrm{i}g \, (\sigma_z+\mathrm{i}\sigma_y)} = 1-\mathrm{i}g \, (\sigma_z+\mathrm{i}\sigma_y),
\end{equation}
and we easily get:
\begin{equation}
    U(q) = \begin{pmatrix}
e^{-\mathrm{i}\lambda}(1-\mathrm{i}g) & -\mathrm{i}e^{\mathrm{i}\lambda}g\\
\mathrm{i}e^{-\mathrm{i}\lambda}g & e^{\mathrm{i}\lambda}(1+\mathrm{i}g).
\end{pmatrix}
\end{equation}
From the perspective of dynamical instability, we can extract the "Lyapunov" exponents by solving the eigenvalue equation 
\begin{eqnarray}
\label{floquet_eqn}
U(q)|\phi_\pm(q)\rangle=e^{i\omega_\pm(q)}|\phi_\pm(q)\rangle
\end{eqnarray} 
with eigenvalues given by
\begin{eqnarray}
\label{quasi-energy}
e^{i\omega_\pm(q)}=(\cos\lambda-g\sin\lambda)\pm\sqrt{g^2-(g\cos\lambda+\sin\lambda)^2}.
\end{eqnarray} 
The appearance of negative imaginary parts in $\omega_\pm(q)$ indicates a dynamical instability, i.e. an exponential growth of the associated mode $q$ at a rate $s_q=-{\rm Im}\omega(q)$. Considering different hopping amplitude $J$ with a fixed weak nonlinearity strength $g\ll 1$, we find that the initial solution with $k_0=0$ is unstable against perturbations when $J+g>\pi/2$, as presented in Fig.\ref{fig:dispersion}b. This instability will grow and dominate the dynamics as soon as the unstable $q$ modes get populated. This explains the reason why we have chosen hopping amplitudes satisfying $J+g<\pi/2$ in the main text, for example $J=1$ or $J=1.3$ for $g=0.1$. It has also been shown that in the prethermal plateau reached by the system, the boiling time is shorter for larger hopping amplitude~\cite{Haldar2021}. It is thus important that our numerical simulation time is larger than the boiling times observed in the prethermal phase. Otherwise, it would be difficult to relate the observed stability of the superfluid state to a genuine robustness against Floquet heating or to a boiling time well beyond the simulation time.

When $\omega(q)$ is real, its associated mode is norm-conserving and stationary. These stroboscopic states are similar to the eigenstates of a time-independent Hamiltonian. To illustrate this point further, we define the Floquet Hamiltonian associated to the first Brillouin zone of the excitation spectrum through:
\begin{eqnarray}
H_F(q) = -\mathrm{i}\ln U(q).
\end{eqnarray} 
It effectively describes the dynamics of the excitations in our periodically-driven system. Generally, this operator can be highly non-local since $U(q)$ features the concatenation of 2 non-commuting unitary operators. There are several methods to approximate $H_F(q)$ in the fast-driving frequency regime~\cite{Goldman2014prx,Bukov2015aip}. Here, we employ instead the Baker-Campbell-Hausdorff (BCH) formula~\cite{Achilles2012bch}. We focus here on the dynamics of low-energy excitations around an initial plane wave state at $k_0=0$, i.e. for $q$ small. In this case, $\lambda \approx Jq^2/2 \ll 1$. We further assume weak interaction strengths $g\ll 1$.

Given two possibly non commuting operators $X$ and $Y$, the BCH formula gives the operator $Z$ satisfying
\begin{eqnarray}
e^{Z} = e^{X}e^{Y}
\end{eqnarray}
as a formal series in terms of $X$, $Y$ and their iterated commutators:
\begin{eqnarray}
\label{eq:BCH}
Z&=&X+Y+{\frac{1}{2}}[X,Y]\nonumber\\
&+&{\frac{1}{12}}[X,[X,Y]]-{\frac{1}{12}}[Y,[X,Y]]+(\cdots)
\end{eqnarray}
where $(\cdots)$ indicates terms involving higher commutators of $X$ and $Y$. Note that the series is not necessarily convergent for generic cases. Plugging $X=-\mathrm{i} \lambda \, \sigma_z$ and $Y=-\mathrm{i} g \, (\sigma_z+\mathrm{i}\sigma_y)$ in Eq.\eqref{eq:BCH}, we can write the first few terms for the effective Floquet Hamiltonian,
\begin{eqnarray}
H_F(q)&=& H^{(1)}_F(q)+H^{(2)}_F(q)+(\cdots) \nonumber\\
H^{(1)}_F(q) &=& \mathrm{i}g\sigma_y + (\lambda + g) \sigma_z \\
H^{(2)}_F(q) &=& -\mathrm{i}g\lambda \sigma_x
\nonumber
\end{eqnarray}
where $(\cdots)$ denotes higher order terms in $g$ and $\lambda$. In the main text, we presented the effective quasi-energies by considering $H^{(1)}_F(q)$ only. Here we consider the quasi-energies $\pm\omega(q)$ at next order, that is for $H^{(1)}_F(q)+H^{(2)}_F(q)$. We find:
\begin{eqnarray}
\omega(q) = \sqrt{gJ} \, |q| \, \sqrt{\frac{J(1-g^2)q^2}{4g}+1 }.
\end{eqnarray}
We observe that the linear spectrum observed at $q\to 0$ for $H^{(1)}_F(q)$ still persists for $H^{(1)}_F(q)+H^{(2)}_F(q)$ with the same sound velocity $c=\sqrt{gJ}$. We also see  that the healing length is slightly renormalised to $\xi = \sqrt{J(1-g^2)/g} \approx \sqrt{J/g}$ since $g\ll 1$. Alternatively, we can numerically diagonalise Eq.\eqref{floquet_eqn} to obtain $\omega(q)$ to all orders and extract the sound velocity from the $q\to 0$ linear behavior.

In Fig.\ref{fig:dispersion}a, we plot $\omega(q)$ as a function of $q$ and a linear dispersion relation is clearly observed for small $q$ up to momenta of order $\xi^{-1}$ with sound velocity $c=\sqrt{gJ}$. Note that the dynamical instability for large hopping amplitude $J$ does not appear in the approximation presented above, since it happens in a regime where the assumption $\lambda \ll 1$ does not hold anymore and originates from higher order terms in the series expansion. As demonstrated in Fig.\ref{fig:dispersion}b, the instability indeed occurs at $q\approx \pi$, i.e. at $\lambda\approx 2J$. An intuitive picture would be that the excitations, coming in pairs $\pm q$ of momenta, proliferate when the excitation gap is on resonance with the driving frequency.

\section{Differences between the superfluid and prethermal phases close to the critical line for the single-site impurity case}
In the phase diagram of Fig.2 of the main text, we concluded that there is a critical line separating a prethermal regime where the Floquet system eventually heats up to infinite temperature from a superfluid regime where no heating ever arises. Here we will present more details on the different dynamical behaviors of the system when it is slightly above or below the critical line.

\begin{figure}
\includegraphics[width=1.0\columnwidth]{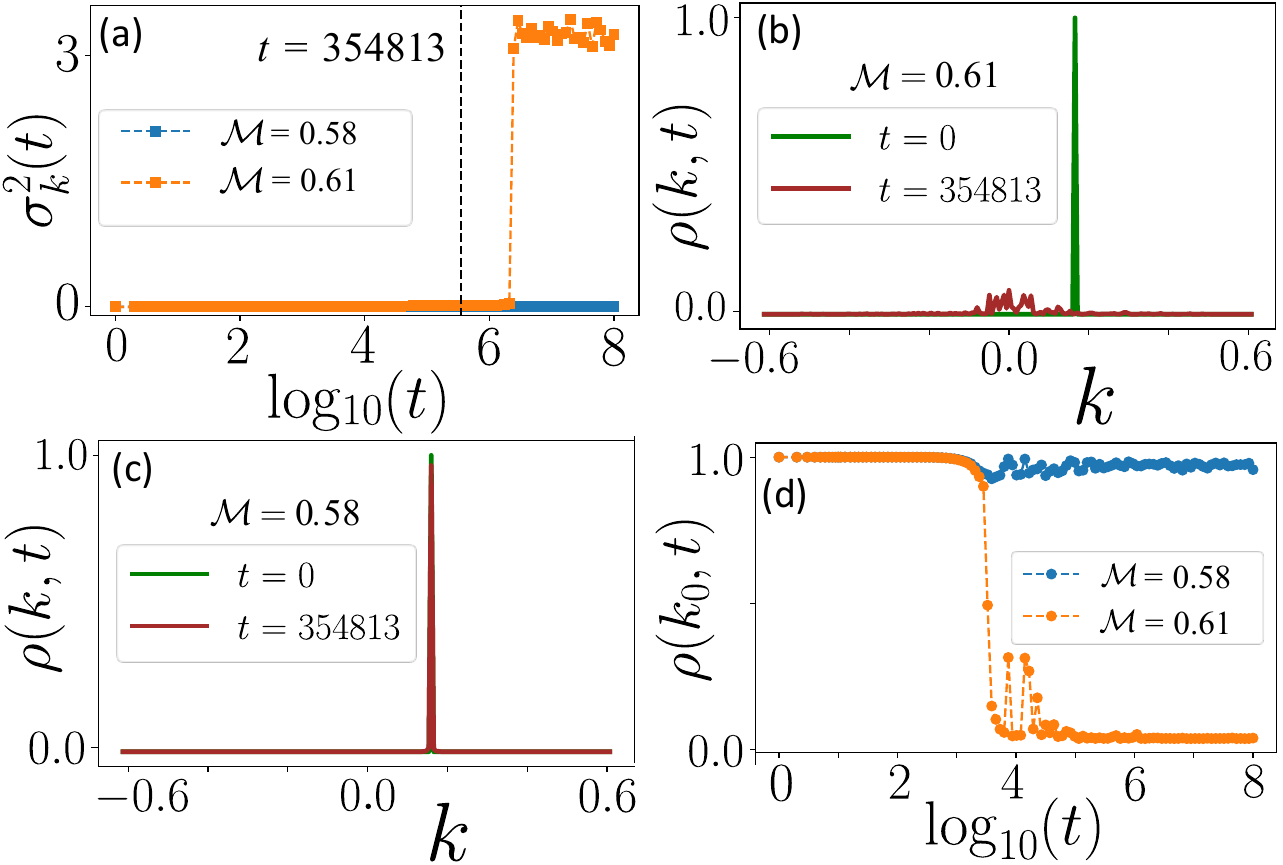}
\caption{Dynamical behavior of our periodicially-kicked system in the presence of a single-site impurity. (a) Momentum variance as a function of time for an initial plane wave state $k_0$ just above (prethermal state) and just below (superfluid state) the critical line. (b) Momentum distribution of the system in the prethermal phase at time $t=0$ and at time $t=354813$ slightly before before $\tau_{\rm boil}$, see dashed vertical line in Panel (a). (c) Same as (b) but in the superfluid phase. (d) Time evolution of the population in mode $k_0$ just above (prethermal state) and just below (superfluid state) the critical line. In all panels, the blue color denotes the superfluid phase and the orange color the prethermal phase. Other parameters values are the same as in Fig. 3 of the main text.
}
\label{fig:diff_rho}
\end{figure}

As seen in Fig.\ref{fig:diff_rho}a, the respective variances $\sigma_k^2(t)$ obtained for an initial mode $k_0$ started slightly above or slightly below the critical line remain identically small for a very long time. As a consequence, they cannot accurately discriminate the two phases, at least until the boiling time $\tau_{boil}$ for the prethermal phase. However, the underlying dynamical behaviors in these two phases are markedly different. This is exemplified in Fig.\ref{fig:diff_rho}b and Fig.\ref{fig:diff_rho}c, where we show the momentum density profiles of our system in the prethermal and superfluid phases at the initial time and at a later time below $\tau_{boil}$ shown by the vertical dashed line in Panel (a). In the prethermal phase close to the critical line, the momentum distribution obtained for initial modes at $k_0\neq0$ shifts towards the "lowest" quasi-energy mode $k=0$ where it grows a momentum component. This phenomenon has been studied in~\cite{Haldar2021} for the disordered case and signals a (classical) wave condensation process at work: the momentum distribution develops a macroscopically occupied mode on top of a thermal distribution described by the Rayleigh-Jeans (RJ) thermal distribution. Importantly, the macroscopically occupied mode is no longer the initial mode. In marked contrast, the momentum distribution obtained in the superfluid phase close to the critical line for initial modes at $k_0\neq0$ remains unaffected. This is further illustrated in Fig.\ref{fig:diff_rho}d where we show the population in the initial mode for both phases as a function of time. The initial mode $k_0$ always remains macroscopically occupied in the superfluid phase while it drops much earlier than the boiling time $\tau_{boil}$ in the prethermal phase.

\section{Derivation of the effective impurity strength in the disordered case}

\begin{figure}
\includegraphics[width=.98\columnwidth]{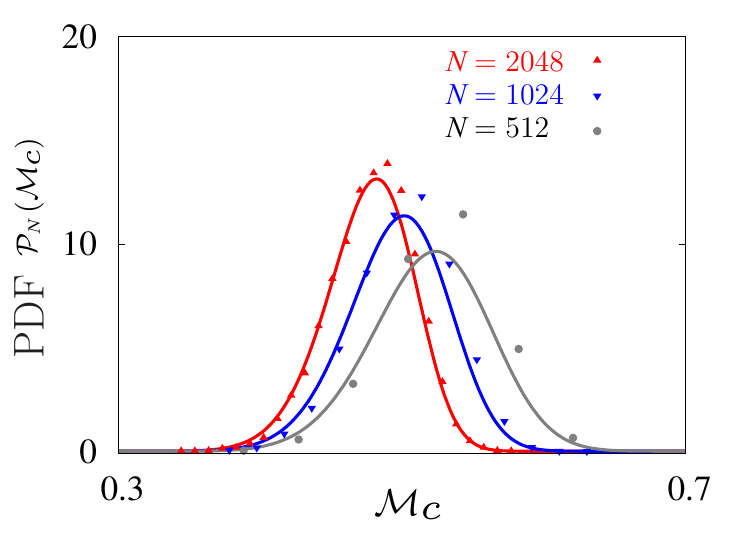}
\caption{Probability distribution function (PDF) $\mathcal P_N(\mathcal M_c)$ of the critical Mach number $\mathcal{M}_c=k_c\xi$ for different system sizes $N$ at constant effective disorder strength $\Lambda=W/c=0.3$ ($J=1$ and $g=0.1$): $N=512$ (grey dots), $N=1024$ (blue dots), and $N=2048$ (red dots). The full curves are analytical predictions extracted from Eq.\eqref{eq_xc_disorder} and symbols are numerical simulations of the disordered Floquet model as explained in the main text. We used $10^4$ disorder realisations.
}
\label{fig:vc_L}
\end{figure}

We explain here the renormalisation procedure that we use to effectively replace the full disorder potential by a single-site impurity model. The first step is a screening argument. Since the system cannot resolve spatial potential scales smaller than the healing length, we first construct a coarse-grained potential by summing up all the impurities in a window of size $\alpha \xi$, with $\alpha$ a parameter of order unity that will be fixed by fitting a single set of data. Starting from $N$ uncorrelated impurities of strength $V_x\in[-\frac{W}{2},\frac{W}{2}]$ with box uniform distribution ($\langle V_x\rangle=0$, $\langle V_x V_{x'}\rangle=\frac{W^2}{12}\delta_{xx'}$), we define a new renormalised potential
\begin{equation}
  U_j=\sum_{x\in \textrm{box }j} V_x.
\end{equation}
Each box contains $M_\alpha=\alpha \xi/a$ impurities, where $a$ is the lattice spacing ($a=1$ in our work) and the total number of boxes is $B=N/M_\alpha$. The idea is then to look for the dominant contribution that comes from the maximum $U_m=\textrm{max}\left\{ U_j \right\}$ of this new disorder coarse-grained potential. At this stage, everything boils down to a problem of extreme value statistics.

We first look at the statistics of the random potential $U_j$. Note that $U_j$ is a sum of independent random variables with the same uniform distribution. It rapidely converges ($M_\alpha\ge 5$ is enough for our purpose) to a normal distribution with zero mean and variance $\sigma^2=M_\alpha \frac{W^2}{12}$. The statistics of the maximum value $U_m= \textrm{max}\left\{ \tilde u_j \right\}$ is a standard problem in extreme value statistics, and we briefly explain it here. It is easy to find the cumulative distribution of the maximum, i. e. the probability for $U_m$ to be smaller than a given value $U$. Indeed, for the maximum to be smaller than $U$, all of the $V_x$ contributing to $U_j$ have to be smaller than $U$. Since the $B$ random variables $U_j$ are independent, the result is just the product of their respective cumulative distributions $F(U)$. We therefore end up with a simple equation for the cumulative distribution $\mathcal F_B$ of $U_m$,
\begin{equation}
  \mathcal F_B(U_m)=\big[F(U_m)\big]^B.
\end{equation}
Note that for a Gaussian variable $z$ with distribution $p(z)=\exp[-z^2/2\sigma^2]/\sqrt{2\pi\sigma^2}$, we have

\begin{equation}
F(z)=\int_{-\infty}^z dy p(y)=\frac{1+\textrm{erf}\left(\frac{z}{\sqrt{2}\sigma}\right)}{2}.
\end{equation}

We now use the crude approximation of replacing the full complexity of the microscopic disorder with a single-site impurity of effective strength $U_m$ and apply the criterion in Eq.\eqref{eq:vc_delta} above to compute the critical Mach number $\mathcal M_c=k_c\xi$, namely
\begin{equation}
  \frac{U_m}{c} = \Lambda_c( \mathcal M_c).
\end{equation}
This relation being one-to-one we finally obtain the statistical distribution of $\mathcal M_c$ through the following relation between the cumulative distributions \begin{equation}\label{eq_xc_disorder}
  \Phi_N(\mathcal M_c)=1-\mathcal F_B(U_m=c \Lambda_c(\mathcal M_c)).
\end{equation}
This yields Eq.(8) in the main text. This simple expression contains however nontrivial scaling properties of $\mathcal M_c$ (and thus $k_c$) with the system size $N$, the disorder strength $W$, and the hopping amplitude $J$ and interaction strength $g$ (via $c$ and $\xi$). Finally, the parameter $\alpha$ is determined by fitting the theory to a single set of data. We obtained $\alpha=2.78$ which is indeed a parameter of order unity.

In addition to the data shown in the main text, we demonstrate in Fig.\ref{fig:vc_L} that the scaling of the critical momentum $k_c$ with respect to the system size is also well reproduced by this model. As the system size $N$ is increased, the distribution is shifted to lower momenta. This can be easily understood: By increasing $N$, we increase the number $B$ of boxes and thus also increase the probability to find a large $U_j$ value. Note however, that the Gaussian approximation for the random variable $U_j$ predicts that the critical momentum $k_c$ vanishes in the thermodynamic limit $N \to\infty$ at fixed number of particles per site $N_a/N$. This is actually an artefact of this Gaussian approximation since $U_j$ is bounded by $M_\alpha W/2$. This pathology can be corrected by taking the true distribution of $U_j$ that can be exactly computed. However, we chose to use the Gaussian approximation in order to avoid cumbersome expressions and calculations that are not really important for general illustration purposes. 

\begin{figure}
\includegraphics[width=.98\columnwidth]{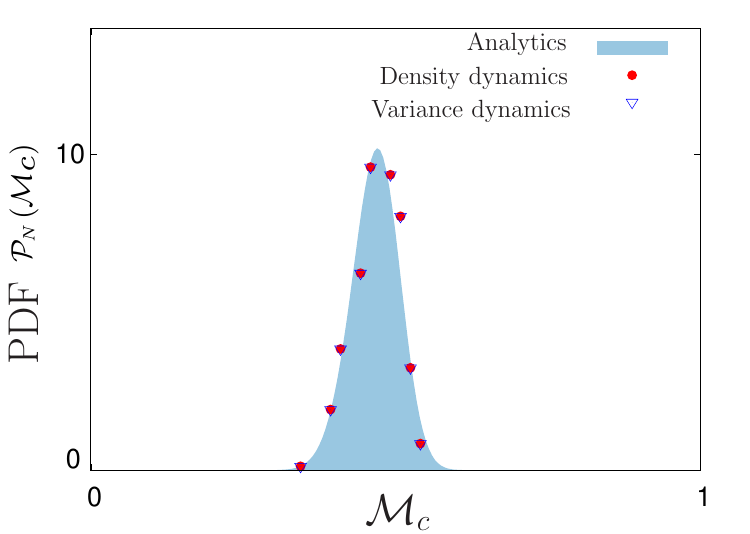}
\caption{Probability distribution function (PDF) $\mathcal P_N(\mathcal M_c)$ of the critical Mach number $\mathcal{M}_c= k_c\xi$ determined from both the dynamical behaviors of the population $\rho(k_0,t)$ (red dots) and of the variance $\sigma^2_k(t)$ (blue triangles) at constant disorder strength $\Lambda=W/c=0.3$ ($J=1.3$ and $g=0.1$). The blue area is the analytical prediction. We have used here $360$ disorder realizations to obtain the PDF.
}
\label{fig:vc_disorder_heating}
\end{figure}

Last, in Fig.\ref{fig:vc_disorder_heating}, we present additional results on the distribution of the critical Mach number separating a finite boiling time and an "infinite" one for the periodically-kicked nonlinear disordered lattice. Note that by "infinite" we mean that we were not able to observe any boiling process for an initial momentum $k_0$ slightly below the critical line up to the longest time explored in our numerical simulation ($10^7$ periods). This longest simulation time was however enough to observe the finite boiling time $\tau_{boil} \sim 10^6$ for $k_0$ slightly above the critical line. We find an excellent agreement between the probability distribution functions (PDF) obtained from the dynamical behaviors of the population $\rho(k_0,t)$ and of the variance $\sigma^2_k(t)$. Specifically, the critical $\mathcal{M}_c$ is obtained as the largest value of $k_0\xi$ for which $\rho(k_0,t)$ never drops below $0.5$ and $\sigma^2_k(t)$ never increases above $0.5$ within the simulation time $10^7$ in each disorder realization. This confirms again that the superfluid phase we report in this Letter for our weakly disordered Floquet model is robust against heating when $J+g<\pi/2$.  

\bibliographystyle{eplbib.bst}